\documentclass[a4paper,11pt]{article}
\pdfoutput=1 

\usepackage{jcappub} 

\usepackage[T1]{fontenc} 

\usepackage{color}

\title{\boldmath Supernovae anisotropy power spectrum}

\author[a]{	Hoda Ghodsi,}
\author[a]{Shant Baghram,}
\author[b]{Farhang Habibi}

\affiliation[a]{ Department of Physics, Sharif University of Technology, P.~O.~Box 11155-9161, Tehran, Iran}
\affiliation[b]{ LAL-IN2P3/CNRS, BP 34, 91898 Orsay Cedex, France}

\emailAdd{h.ghodsi@mehr.sharif.ir}
\emailAdd{baghram@sharif.edu}
\emailAdd{habibi@lal.in2p3.fr}

\abstract{We contribute another anisotropy study to this field of research using Type Ia supernovae (SNe Ia). In this work, we utilise the power spectrum calculation method and apply it to both the current SNe Ia data and simulation. Using the Union2.1 data set at all redshifts, we compare the spectrum of the residuals of the observed distance moduli to that expected from an isotropic universe affected by the Union2.1 observational uncertainties at low multipoles. Through this comparison we find a dipolar anisotropy with tension of less that $2\sigma$ towards $l =171^\circ \pm 21^\circ$ and $b=-26^\circ \pm 28^\circ$ which is mainly induced by anisotropic spatial distribution of the SNe with $z > 0.2$ rather than being a cosmic effect. Furthermore, we find a tension of $\sim 4\sigma$ at $\ell=4$ between the two spectra. Our simulations are constructed with the characteristics of the upcoming surveys like the Large Synoptic Survey Telescope (LSST), which shall bring us the largest SNe Ia collection to date. We make predictions for the amplitude of a possible dipolar anisotropy that would be detectable by future SNe Ia surveys.}

\begin{document}
\maketitle
\flushbottom

\section{Introduction}
\label{sec:intro}

The observational investigation of the validity of the cosmological principle (CP) and its domain of validity is one of the main questions of modern cosmology. The isotropy of the cosmic microwave background radiation (CMB) suggests that the Universe is isotropic on very large scales (of order 100$h^{-1}$Mpc). Despite this fact, we observe cosmic structures such as voids and super-clusters in the nearby Universe. One then wonders where the transition between these two states occurs. While we have data on very large scales and also in our neighbourhood, we still require further data on the intermediate scales. This is where Type Ia supernovae (SNe Ia), which are the subject of our study in this work, could come useful. On the other hand, beside the important question of checking the validity of the CP, there are some known anomalies which are related to the possible anisotropy of the Universe. Accordingly, the search for a possible preferred axis, and hence anisotropy, in the cosmos has more critical motivations than to just look for a transition scale of the CP. These are as follows: 

\begin{itemize}
\item \textit{Large scale velocity flows}: The scale of large scale (of order $100^{-1}$Mpc or larger) bulk flows is observed to be greater than what is expected in the standard $\Lambda$CDM cosmology \cite{Kash,Watkins,Lavaux}.

\item \textit{The alignment of the CMB power spectrum low multipoles}: The directions of the normals to planes of the octopole and quadrupole moments and the dipole moments in the observed CMB map seem to point to a unified direction \cite{Eriksen,Schwa}.
\item \textit{The CMB power asymmetry}: There is an indication of a power asymmetry in WMAP \cite{Eriksen:2007pc} and Planck data\cite{Ade:2013nlj}.

\item \textit{Large scale alignment of quasar optical polarisation data}: It turns out that the quasar polarisation vectors point towards a common direction in the sky \cite{Hutse1,Hutse2}.
\end{itemize}

\vspace{2mm}
If there is indeed anisotropy discovered in the Universe which is trustable and not due to systematic effects, various proposed physical effects could be responsible for such a signal. First and foremost the founding assumption upon which the standard cosmological paradigm is constructed would not hold any longer. Another possibility would be a dark energy with an anisotropic equation of state \cite{Koivis}. And also early universe models can introduce anisotropies \cite{Erickcek,Abolhasani}. In using SNe Ia as our anisotropy probe, there could occur various events which could disguise as anisotropy signals. Such these effects are the following: The intrinsic scatter of the SNe Ia (due to them not being perfect standard candles), scatter due to the location of the SNe Ia within the host galaxy and the type of the galaxy, extinction due to dust in the host galaxy, intergalactic medium and our own galaxy and finally, gravitational lensing along the line of sight to the SNe Ia, which could alter the light coming from the source \cite{Aman,Baghram}. But since most of the effects mentioned happen on the galactic scale they will be averaged out in a statistically large enough sample \cite{Kolatt}.

Various methods have been employed over the years for investigating possible anisotropies in the Universe at varying scales. These are either model independent searches or ones assuming certain anisotropic models \cite{Alnes,Jain,Cai}, for a specific example we can mention dark energy models \cite{Koivisto,Blomqvist1,Blomqvist2,Camp}. With regards to the model independent approaches, many works employ the hemisphere comparison method, like \cite{Schwarz,Anton,Cai2,Tsagas,Kalus,Javan}. Other works such as that of Colin et al.'s in \cite {Colin} employ the statistical tool of `residual' they developed and analyse the data tomographically. Also the dipole anisotropy is studied \cite{Cooke,Lin} as well. 

All of these works find low significance anisotropies, which are mostly attributed either to systematic effects or low numbers of SNe Ia. While obviously these two sources of uncertainty are to be rectified in different ways, it is noteworthy to point out that as the number of data points (SNe Ia) increases in future data sets, the sources of systematics will be identified more clearly and therefore we will know how to try and improve upon our methodology and/or technologies we utilise at the moment.

In this work we use the method of angular power spectrum calculation and apply it to the deviation of SNe Ia distance moduli from those in the standard isotropic model in order to probe anisotropy in the dipole and higher multipole moments. We also demonstrate that this method is potentially a strong probe to investigate anisotropy with future larger data sets.

We should also mention that there exist studies that are pursued in the same direction \cite{Kolatt,Zhao,Carvalho}. One such important study is the work of Bengaly et al. \cite{Bengaly}. These authors consider only the low-redshift regime in order for their analysis to be model independent. Making use of the two most recent SNe Ia data sets available namely the Union2.1 and the JLA data set \cite{Betoule}, Bengaly et al. conclude that they cannot discard the possibility of the existence of a genuine anisotropy in the recent Universe. We point out here that in the present work, in contrast to recent similar studies, we probe the angular power spectrum of SNe Ia at higher redshifts, mainly with the motivation of bulk flow measurement of possibly cosmological origin. This would allow us to probe any possible discrepancy with the standard cosmological model prediction as discussed in the recent work by Colin et al. \cite{Colin:2017juj}. We should therefore observe some degree of dipolar anisotropy in the SNe Ia data, which is also found in many of the previous works albeit at low significance \cite{Colin,Schwarz,Anton}.

Also all of the power spectrum investigations cited above utilise the SNe Ia data sets only, whereas we use the data as well as simulations in order to try and make predictions regarding what future data could tell us in this regard. We aim to determine a threshold value for the amplitude of a possible dipole anisotropy seeable by future supernova surveys such as the LSST \cite{LSST}. 

The paper is organized as follows. In section 2 we describe the methods we employed in our investigations and analyses. In section 3 we introduce the observational data we made use of in our work and investigate any detectable anisotropies out of the data. In section 4 we talk about our simulations for future surveys including dipolar anisotropy. Lastly, in section 5 we shall conclude.

\section{Method}

In this section the method used in our search for finding possible anisotropies in the Universe is explained. As we mentioned briefly in the introduction, our method is that of power spectrum calculation. In this method, we simply target every supernova in turn and calculate their theoretical distance moduli to compare with their observed values. The distance modulus is given as:

\begin{equation}
\mu = 5 \log d_L + 25,
\label{mumod}
\end{equation}

where $d_L$ is the luminosity distance in Mpc given in the standard cosmological model ($\Lambda$CDM) by:

\begin{equation}
d_L = \frac{c(1+z)}{H_0}\int_{0}^{z} \frac{dz'}{\sqrt{\Omega_M (1+z')^3 + \Omega_{\Lambda}}},
\end{equation}

where the integral is from now $(z=0)$ to the redshift of the object, $z$, $c$ is the speed of light, $\Omega_{M}$ and $\Omega_{\Lambda}$ are the matter and cosmological constant density parameters and $H_0$ is the Hubble parameter of expansion. Note that we have assumed a flat Universe with zero curvature.
Now with the theoretical calculation done we can build a field of distance modulus residuals, $\Delta \mu$, of the differences between our model predicted distance moduli, $\mu_{\rm mod}$ and the observed ones, $\mu_{\rm obs}$:

\begin{equation}
\Delta \mu=\mu_{\rm obs}-\mu_{\rm mod}.
\end{equation}

We take as our fiducial model, the standard $\Lambda$CDM model with the parameters taken from the Union2.1 SNe Ia data \cite{Suzuki} to be consistent in our supernova investigation. Assuming a flat universe, we take the matter density parameter to be $\Omega_{M} = 0.295 \pm 0.034$ and we fix the Hubble constant value at, $H_0 = 70$ km/s/Mpc.

\vspace{2mm}

The residual field created is now expanded in spherical harmonics:

\begin{equation}
\Delta \mu=\sum_{\ell=0}^{\ell=\ell_{\rm max}} \sum_{m=-\ell}^{m=+\ell} a_\ell^m Y_\ell^m.
\end{equation}

And the power spectrum is given as:
\begin{eqnarray}
C_\ell &=& \frac{1}{2\ell+1}\sum_{m=-\ell}^{+\ell} a_{\ell m} a_{\ell m}^*.
\label{cl}
\end{eqnarray}

In the next section we will use the Union2.1 data set and apply the power spectrum calculations as defined above to this data set.


\begin{figure}[ht!]
\centering
\includegraphics[scale=0.2]{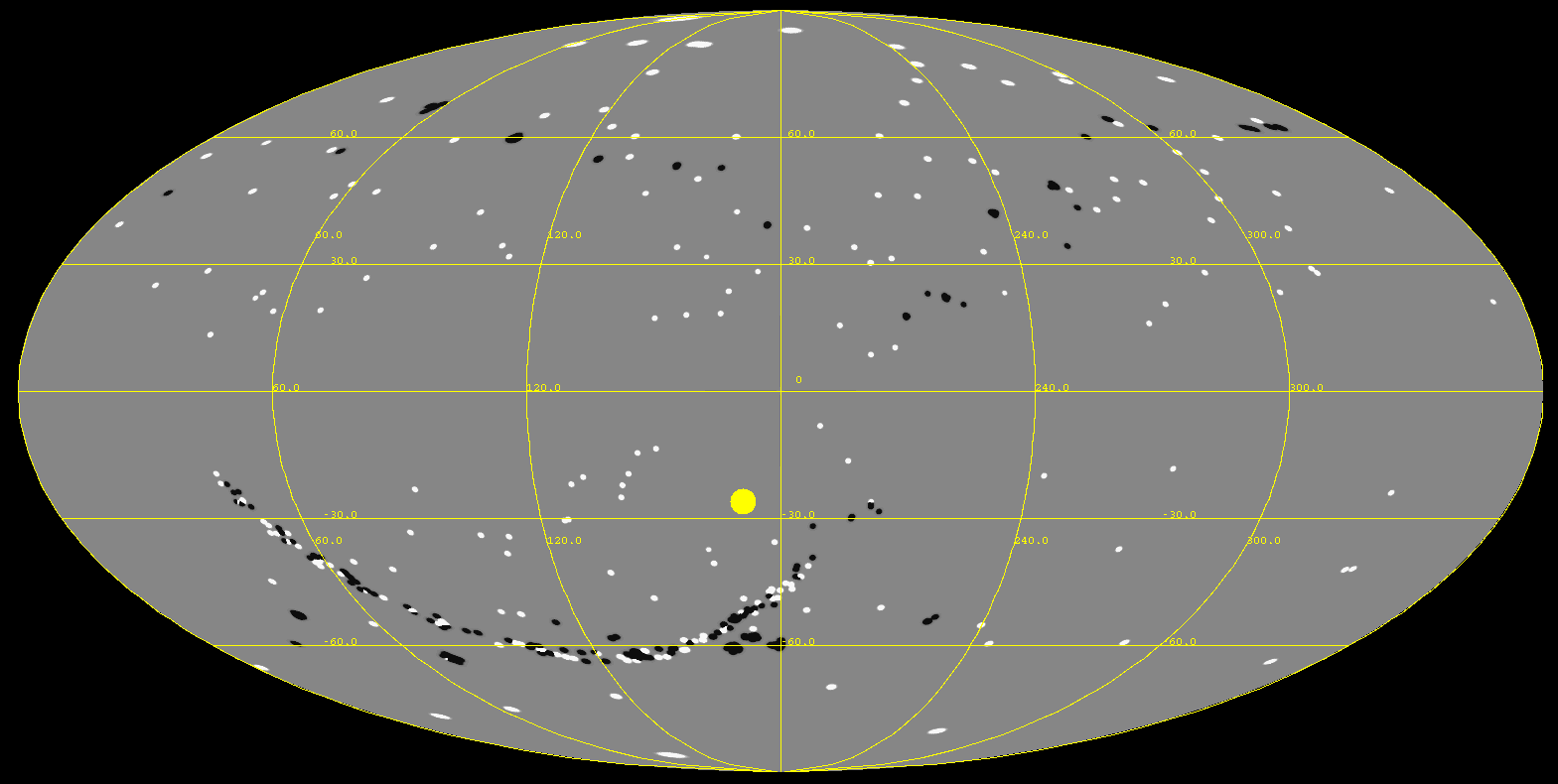}
\caption{The distribution of the Union2.1 SNe Ia in Galactic coordinates. The white and black points represent SNe with redshifts smaller and larger than 0.2 respectively. The yellow full circle shows the direction of the detected dipole as discussed in section 3.1.}
\label{u21}
\end{figure}

\section{Observational Data}

We utilise the Union2.1 SNe Ia data set \cite{Suzuki} in our study. The 580 supernovae in this data set span a redshift range of $0.015<z<1.414$ and the sky distribution of this data set is shown in figure \ref{u21}. This data set is put together by combining various data sets after analysis through the unified light curve-fitter of SALT2 \cite{Guy} and it follows the previous Union \cite{Kowalski} and Union2 \cite{Aman} data sets.

\subsection{Results from Union2.1}

To search for any detectable anisotropy in the Union2.1 data, we compare the angular power spectrum ($C_\ell$) of the observed distance modulus residuals, $\Delta \mu$, to the mean spectrum of an isotropic universe which are affected by the Union2.1 observational uncertainties.
For each SN, the distance modulus for the isotropic flat universe, $\mu_{\rm mod}$, is computed for redshifts given by the Union2.1 catalogue and the cosmological constants, $H_0=70$ km/s/Mpc and $\Omega_M=0.295$ \cite{Suzuki} as mentioned in section 2.

According to figure \ref{u21}, the data points are sparse and far from continuity except for the SDSS stripe 82. Expectedly, this will induce fictitious fluctuations on $\Delta \mu$ spectrum specially on small angles. To avoid this, we apply a low-pass angular frequency filter to the data and will consider only the variations on large angular scales (low $\ell$ values).

After calculating $\Delta \mu$ for each SN, we consider a HEALPix spherical map with an angular resolution given by the $N_{\rm side}$ parameter and fill the map pixels with the $\Delta \mu$ values according to the corresponding SNe Galactic coordinates. The pixels containing more than one SN are filled with the mean value of the corresponding residuals.

$a_{\rm \ell m}$ coefficients are then computed from the  $\Delta \mu$ map with a given $\ell_{\rm max}$. By choosing a smoothing angle of $\sigma=13^\circ$, the $a_{\rm \ell m}$'s of both maps are multiplied by $\exp(-\ell(\ell+1) \sigma^2/2)$. This will wash out fluctuations with $\ell>6$. And the power spectrum does not depend on the map resolution for $N_{\rm side} \geq 1024$.

\begin{figure}
\centering
\includegraphics[scale=0.3]{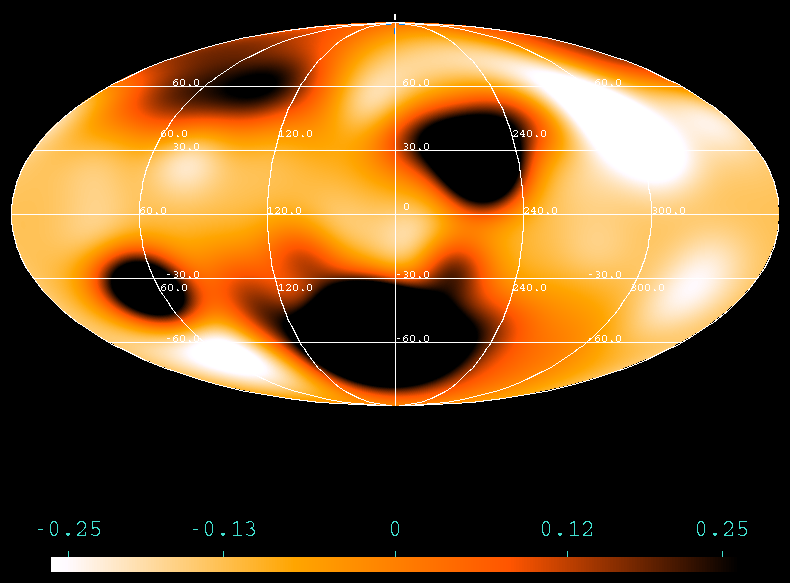}
\caption{The smoothed map of $\Delta \mu$ for SNe Ia of the Union2.1 sample. The smoothing angle is 13$^\circ$. The luminous/dark regions show SNe brighter/dimmer that the prediction for an isotropic universe. The colour scale varies from $-0.25$ to $+0.25$ magnitudes.}
\label{datamap}
\end{figure}

The resulting map is shown in figure \ref{datamap}. The brighter structures show the regions containing SNe with luminosities brighter than the prediction of the isotropic model. And the darker regions represent the dimmer SNe. Using this map, we can compute the associated $C_\ell$'s from eq. \eqref{cl}.

\begin{figure}
\centering
\includegraphics[scale=0.3]{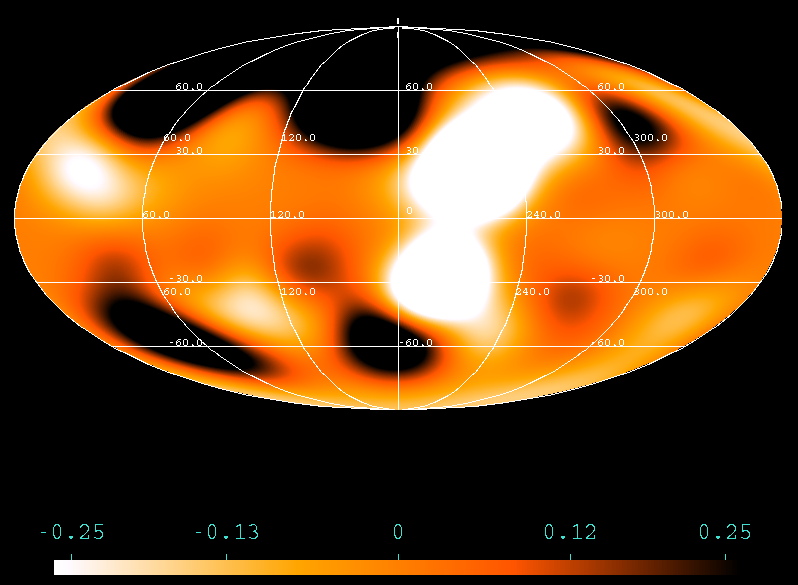}
\caption{Similar to figure \ref{datamap} but constructed for a realisation of an isotropic universe affected by observational uncertainties of the Union2.1 data.
}
\label{datamapiso}
\end{figure}

Now in order to determine the mean power spectrum of the isotropic universe we do as follows: For each Union2.1 supernova, we assume a Gaussian distribution centred at $\mu_{mod}(z)$ with
standard deviation $\sigma_{\rm obs}$ (the observed uncertainty) which includes both statistical and systematic errors. By taking random values for the distance modulus, $\mu_{\rm rnd}$, out of every Gaussian distribution we build up 1000 mock data sets each of which contains 580 SNe.
The residual field for the SNe of every set would then take the form:
\begin{equation}
\Delta \mu=\mu_{\rm mod}-\mu_{\rm rnd}.
\label{mumodsim}
\end{equation}
Figure \ref{datamapiso} shows the smoothed map for residuals computed through relation \eqref{mumodsim}.

With the mock catalogues created, we calculate the corresponding power spectra of residuals
and average over the 1000 realisations to obtain the mean power spectrum with the associated error bars ($1\sigma$ dispersion). It is needless to say that any real anisotropy signal to be detected should surpass these error bars with sufficient significance. In this way we can eliminate possible fake anisotropies.

\begin{figure}
\centering
\includegraphics[scale=0.3]{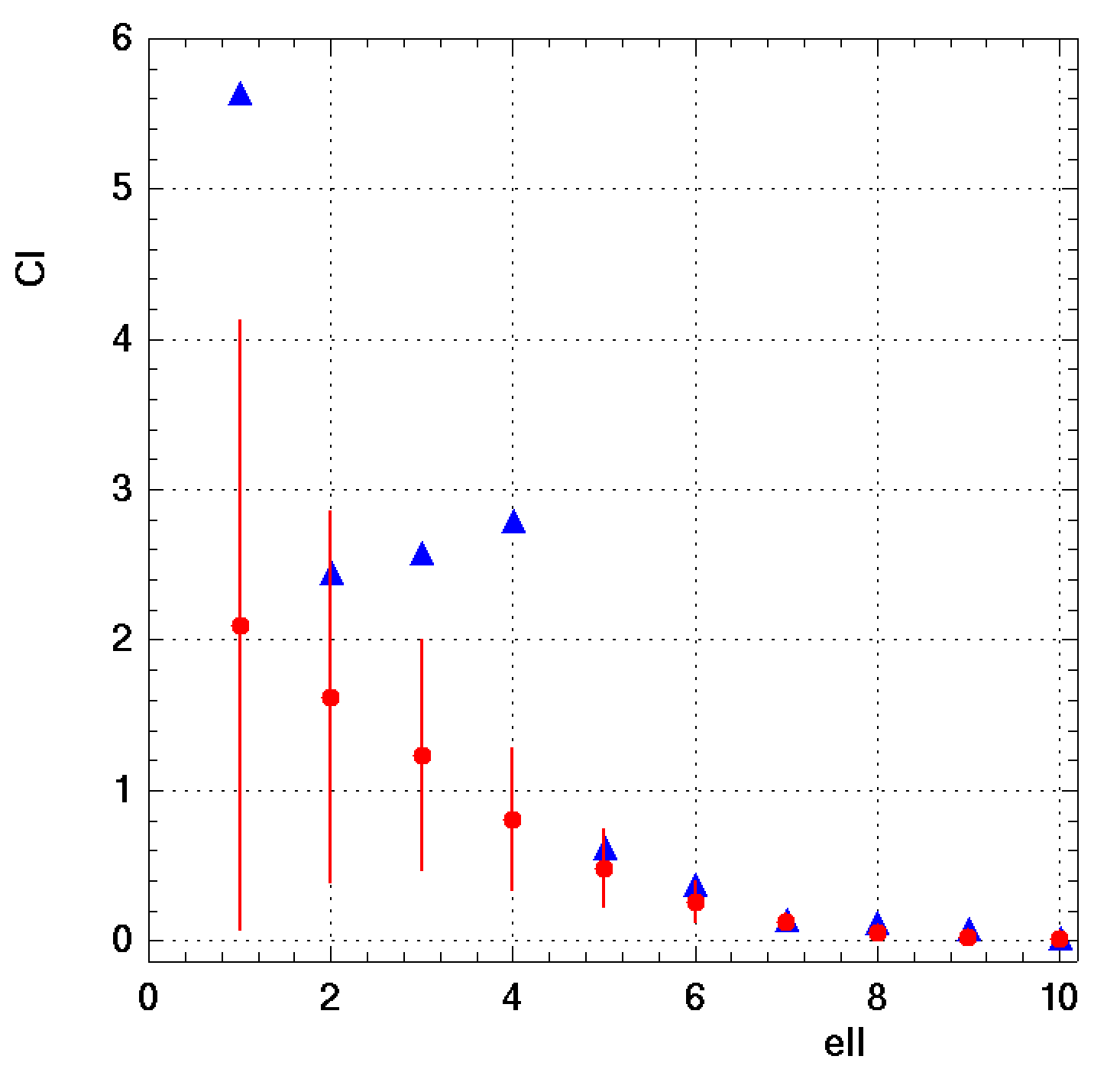}
\caption{The angular power spectrum (per steradian) of the Union2.1 data is shown by blue triangles. The red circles show the mean power spectrum expected from an isotropic universe affected by observational uncertainties of the Union2.1 data.}
\label{spectre}
\end{figure}

The blue triangles in figure \ref{spectre} represent the angular power spectrum of the smoothed data. The peak at $\ell=4$ corresponds to the dark/luminous patterns presented in figure \ref{datamap}. And the red circles in figure \ref{spectre} show the mean power spectrum of the isotropic universe. The error bars represent $1\sigma$ dispersion at either sides of the mean values.
The two spectra in figure \ref{spectre} are separated by $\sim 4\sigma$ at $\ell=4$ with separation less than $2\sigma$ for $\ell=1$. The spectra are derived from maps with $N_{\rm side}=1024$.

The dipole detected locates towards $l =171^\circ \pm 21^\circ$ and $b=-26^\circ \pm 28^\circ$ with an apparent magnitude difference of 4 magnitudes. We should mention that the observed dipolar tension disappears if we redo the analysis excluding  the high redshift SNe with $z>0.2$. According to figure \ref{u21} the observed high redshift SNe Ia with $z > 0.2$ (black points) are distributed less isotropically than the ones at lower redshifts (white points with $z < 0.2$). Indeed, the 230 low redshift SNe Ia are rather uniformly distributed over both hemispheres, 
while $2/3$ of the 350 high redshift SNe Ia are located in the southern Galactic plane, mainly concentrated in the SDSS stripe 82 which is close to the direction of the detected dipole. This indicates that the observed dipole is most likely due to the anisotropy in the observations, rather than being of a cosmological nature. 



\section{Simulation}

In this section, we shall discuss in detail the assumptions made in the production of the simulations we built. We simulated SNe Ia data sets in light of the upcoming more abundant supernovae data. Among these are the LSST \cite{LSST}, Euclid \cite{Euclid} and WFIRST \cite{Wfirst} surveys. As it is presented in the work by Jain et al. \cite{Jain2}, by putting together the results of these future surveys we would be able to achieve much more than if we were to utilise any one of them on their own. For instance, while the LSST will discover SNe Ia in the southern sky, Euclid and WFIRST are planned to explore both hemispheres. Furthermore, putting together the data from all these surveys an overall extended redshift range will be achieved. Of course, to actually put the data from these surveys together requires careful study of the systematics involved in each one but here we are considering creating a simplistic simulation and so suffice to the mention of these facts.

Again in the spirit of creating a simplistic simulation, we assume our data will be sufficiently sky-covering and also spanning the redshift range up to cosmological scales so that we can investigate cosmological (an)isotropy. Furthermore, we require that the precision of the data is good enough such that we can have precise distances to SNe Ia. In this sense we shall adopt the characteristics of the upcoming LSST survey for building our simplistic simulations. For instance regarding the precision of measurements and the systematics involved we shall adhere to the predictions made by the LSST team which we will discuss in section 4.1.

In particular with regards to the issue of the Galactic plane extinction, we neglect to mask it and consider the SNe homogeneously and isotropically distributed over whole sky. It is of course needless to emphasise that we choose an arbitrary anisotropy direction which avoids the Galactic plane.

Moreover, the LSST \cite{LSST} will bring about a vast number of SNe Ia in its 10 years of operation. It is estimated that about 50,000 supernovae per year will be observed using this 6.5 wide telescope. The supernovae range in redshift from 0.1 to 1.

As mentioned previously, we would like to find a threshold value for the amplitude of a possible dipole anisotropy which could be seen by future surveys. For this we constructed two types of isotropic and anisotropic simulations for increasing numbers of SNe Ia. In what follows we explain in detail the methods we used in making these simulations.

\begin{figure}
\centering
\includegraphics[scale=0.3]{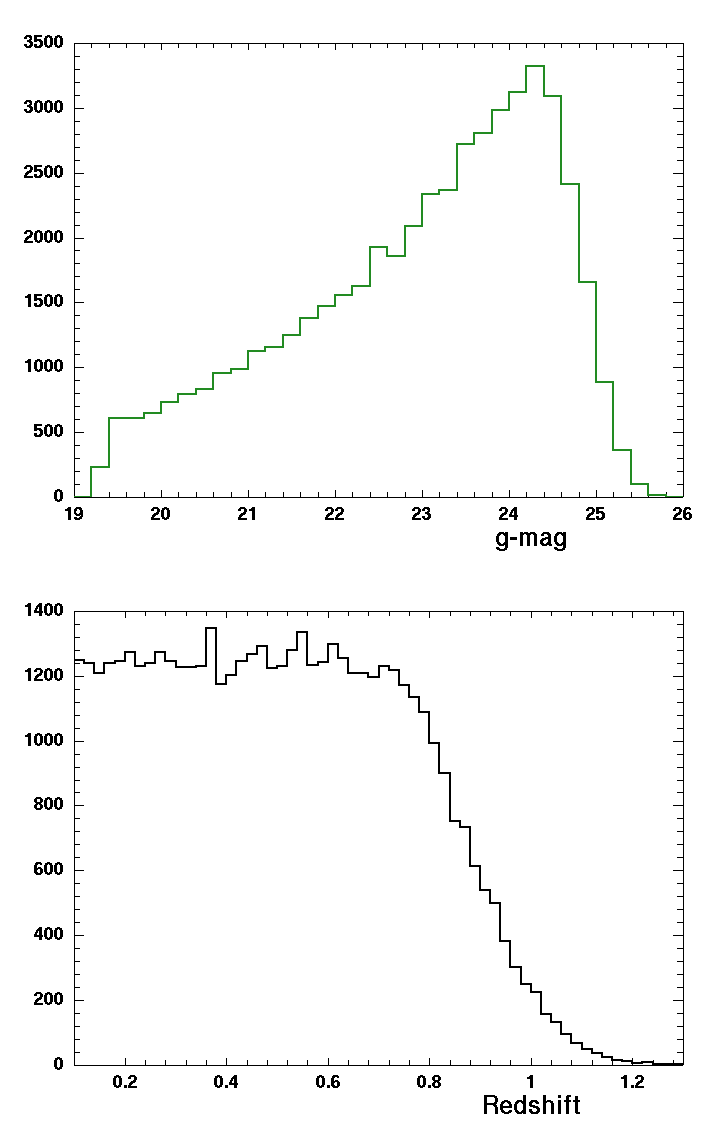}
\caption{The distribution of the simulated apparent g-magnitude of SNe Ia is displayed at the top. The corresponding redshift distribution is shown at the bottom.}
\label{magzdist}
\end{figure}

\subsection{Isotropic Universe Simulation}

To simulate the mean spectra of sky realisations of an isotropic universe, we homogeneously fill the whole sky with 10,000 to 50,000 SNe Ia with a step size of 5,000 and assign isotropic distance moduli to every supernovae according to the random redshifts taken out of the redshift distribution derived from the LSST photometric sensitivity considerations. As for the errors we should include in our analysis, we take into account statistical and systematic errors as expected from the LSST survey which we explain in detail below.

We make 700 sky realisations (mock catalogues) for every number of SNe in turn, in a similar way and average over the resulting power spectra derived from the maps to obtain the mean isotropic power spectrum we are looking for.

To include the statistical errors in the simulated distance moduli, we consider the statistical photometric error predicted for the LSST \cite{LSST}. Considering a supernova with apparent magnitude $m$, the error on the magnitude is given by:
\begin{equation}
\sigma_{\rm stat}^2 = (0.04-\gamma) x + \gamma x^2,
\label{sigstat}
\end{equation}
where $x = 10^{(m-m_5)}$, $\gamma= 0.038$ and $m_5=25$ in g-band. The magnitude limit on the g-band for the LSST single visit is 24.3 \cite{LSST}. We generate SNe Ia located at redshifts larger than 0.1 and compute their distance moduli from eq. \eqref{mumod}.
Neglecting the intrinsic dispersion of SNe Ia, we assume the absolute magnitude for SN Ia in g-band to be $M_g=-19$.

The intrinsic dispersion can be later included as a systematic error. We compute the apparent magnitudes and accept those that pass the magnitude limit of the LSST where the uncertainty on the magnitude limit (given by eq. \eqref{sigstat}) is taken into account. Figure \ref{magzdist} shows the distribution of the g-magnitude and corresponding redshifts for 50,000 SNe which pass the magnitude filter for the LSST single visit. Different realisations of the isotropic universe will follow the magnitude distribution represented in this figure.

For the systematic errors on magnitudes, we consider a simple assumption that the systematic error, $\sigma_{\rm sys}$, is constant for all SNe at all directions. Of course direction-dependent systematics will introduce more complications to detecting anisotropic SNe luminosities, which is beyond the scope of the present study.

Suppose for a realisation of an isotropic universe, we simulate $N$ number of SNe Ia with
apparent magnitudes and redshifts selected from figure \ref{magzdist}. The fluctuation of the distance modulus of the $i$th SN induced by statistical and systematic uncertainties can be quantified as:
\begin{equation}
\Delta \mu^i_{\rm iso} = \delta_{\rm stat}^i + \delta_{\rm sys},
\label{isodelmu}
\end{equation}
where $1\leq i \leq N$. For each SN $\delta_{\rm stat}^i$ is randomly selected from a Gaussian
distribution with a width of $\sigma_{\rm stat}$ and zero mean value. And also $\delta_{\rm sys}$ is constant for all SNe of a realisation changing from one realisation to the next through a random selection from a Gaussian distribution with a width of $\sigma_{\rm sys}$ and zero mean value.

\begin{figure}
\centering
\includegraphics[scale=0.3]{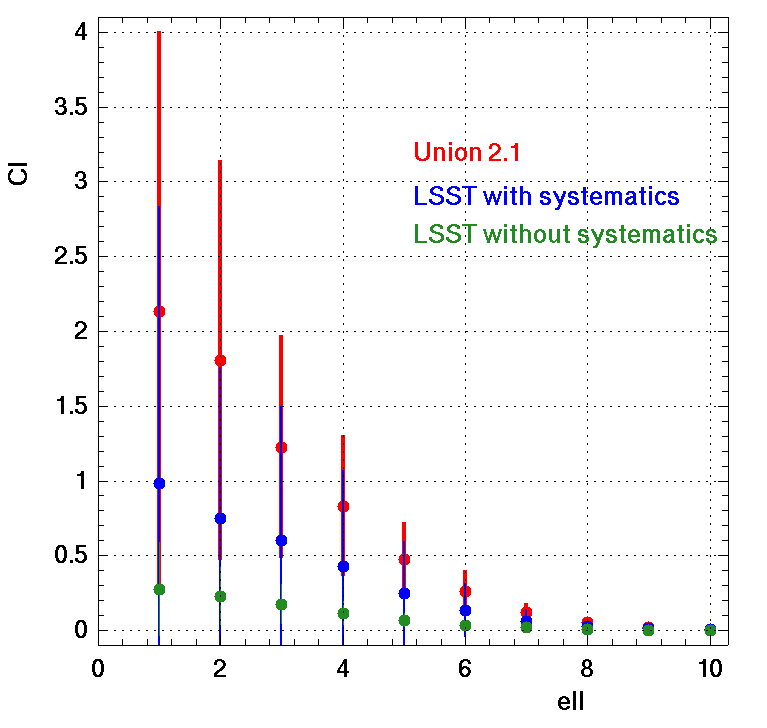}
\caption{The comparison of the mean spectra of isotropic universes estimated using observational uncertainties of the Union2.1 data and the LSST forecast. The error bars indicate $1\sigma$ dispersion.}
\label{isocomp}
\end{figure}

Figure \ref{isocomp} shows the comparison between the spectra computed for the isotropic universe affected by the observational uncertainties of the Union2.1 data and the LSST simulated data. They are produced for the same number of SNe as the Union2.1 data set with the same Galactic coordinates.
Each power spectrum is calculated from the averaging of 700 realisations. The red points are as same as those in figure \ref{spectre} for the Union2.1 catalogue. The green points show the spectrum computed for the LSST including only statistical errors. The blue points include both statistical and systematic errors with $\sigma_{sys}=0.15$. 

At low multipoles, the LSST powers are on average about seven times weaker than those for the Union2.1 in the absence of the systematics. This ratio changes to two in the presence of the systematics where the power dispersion also increases. This shows that a high precision survey (like the LSST) can improve the constraining of the anisotropies up to about an order of magnitude depending on the systematics.    

\begin{figure}
\centering
\includegraphics[scale=0.3]{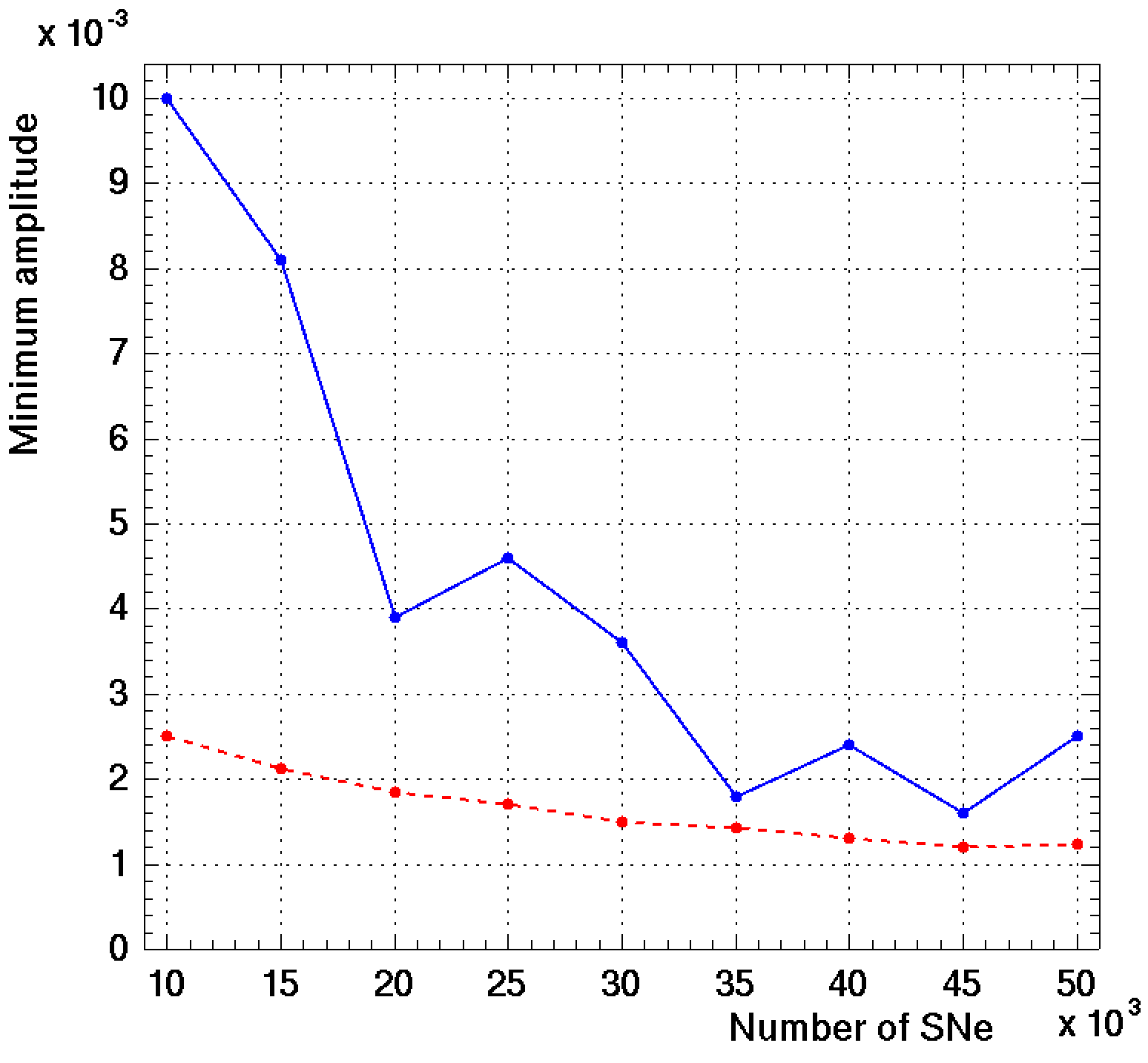}
\caption{Minimum detectable amplitude for a dipole anisotropy as a function of number of SNe Ia distributed homogeneously over the whole sky.
Solid blue curve includes both statistical and systematic errors with $\sigma_{sys}=0.15$. Red dashed curve includes only statistical errors on distance moduli.
}
\label{minA}
\end{figure}

\subsection{Anisotropic Universe Simulation}

The simulation method for an anisotropic universe is similar to that was explained in the previous section only that now we only make one map by choosing an arbitrary direction and applying a modulation to the luminosity distance which results in a modulation in the distance modulus of the form:
\begin{equation}
\mu_{\rm modulated} = \mu_{\rm mod} + 5 \log(1+ A \cos \theta),
\label{anismu}
\end{equation} 

where, $\mu_{\rm modulated}$, is the modulated distance modulus, $\mu_{\rm mod}$, is the model predicted distance modulus, $A$, is the amplitude of the dipole anisotropy and $\theta$ is the angle between the direction of every supernova and an arbitrarily chosen direction of modulation.

Now to have an estimation on the threshold amplitude of a detectable dipole anisotropy described by eq. \eqref{anismu}, again we make a whole sky simulation according to the LSST photometric sensitivity considerations. Once again we uniformly fill the sky from 10,000 to 50,000 SNe Ia with a step size of 5,000 and investigate how the threshold dipole amplitude changes with increasing numbers of SNe Ia in the presence of the statistical and systematic errors. We consider the detection threshold as a minimum amplitude which produces $5\sigma$ separation from the mean spectrum of the isotropic universe at $\ell$=1, where $\sigma$ is the dispersion of the mean spectrum of the isotropic universe as explained in previous section.


\subsection{Results from Simulation}

In this section, we investigated how the increasing number of SNe Ia to be observed in the future can affect the detectable threshold of a possible dipolar anisotropy in SNe Ia luminosity. We determined the amplitude, $A$ of eq. \eqref{anismu} in a way that the anisotropic power spectra would have at least $5\sigma$ separation from the corresponding mean isotropic spectra at $\ell=1$. 

Figure \ref{minA} shows the minimum detectable amplitude for a dipolar anisotropy simulated through the steps explained in section 4.2. The red dashed curve displays the threshold amplitude by including only the statistical uncertainties. And the blue solid curve contains both statistical and systematic errors with $\sigma_{\rm sys}=0.15$. It can be seen that the threshold amplitude decreases by about five times as the number of SNe increases from 10,000 to 50,000 for this curve. Comparing the two plots shows that the presence of the systematics increases the detection threshold amplitude. What is more, we show that future high-precision surveys can improve the anisotropy detection up to about an order of magnitude as compared to the Union2.1 data.

\section{Conclusions}

In this paper we have studied the method of power spectrum calculation in order for identifying possible anisotropies in the Universe using the data of Type Ia supernovae (SNe Ia). For this, we considered using the Union2.1 SNe Ia data set (including all redshifts) and also simulating the future more abundant SNe Ia data of e.g. the LSST.

To analyse the Union2.1 data set, we focused on the spectral values at low multipoles since the data were scarce. Furthermore, as discussed in section 3.1, we adopted a smoothing method which would assist us with our power spectrum calculation for the discontinuous field of our residuals. We then compared our data power spectrum with the equivalent for an isotropic universe. For this we created 1000 sky realisations for the same spatial distribution of the Union2.1 data and computed the mean spectrum of the distance modulus residuals corresponding to an isotropic universe affected by the observational uncertainties of the Union2.1 data. 

Through this comparison, we found $\sim 4\sigma$ tension at $\ell=4$ between the two spectra. An indication of a dipolar anisotropy exists with a significance of smaller than $2\sigma$ towards $l =171^\circ \pm 21^\circ$ and $b=-26^\circ \pm 28^\circ$. We did not detect any significant evidence for dipolar anisotropy at redshifts smaller than 0.2. We found that the observed dipole could hence be an artifact of the anisotropic spatial distribution of the high-redshift SNe Ia rather than being the result of cosmological effects.

We then addressed the anisotropy search for future surveys with the aim of investigating how the increase in the number of observed SNe Ia in the future can affect the detectable threshold of a possible dipolar anisotropy in SNe Ia luminosity. To do so, we considered from 10,000 to 50,000 SNe Ia with a step size of 5000. For each set of SNe Ia, we made 700 realisations of the isotropic universe and computed the corresponding mean angular power spectrum. Then we created anisotropic simulations by adding a dipole term to the distance modulus relation (as shown in eq. \eqref{anismu}) and calculated the resulting anisotropic power spectra corresponding to each SNe Ia set. As for the uncertainties, we used the photometric uncertainties as predicted for the LSST survey in addition to adopting a conservative systematic error of 0.15 magnitudes.

We calculated threshold values with 5$\sigma$ significance for possible dipole anisotropies that could be seen using the power spectrum method for increasing numbers of SNe Ia in the future data sets. As would be expected, such a dipole amplitude decreased in amplitude as the number of observed SNe Ia increased. As discussed at length in section 4, our simulations were simplistic versions of a proper one which would take into account every factor that could possibly affect the final results. As such we did not mask the Galactic plane and therefore, we can say that the dipole amplitudes we have calculated in this investigation would correspond to every direction other than the those towards the Galactic plane.

Moreover, we compared the isotropic power spectra affected by the Union2.1 data observational uncertainties to the one with the predicted uncertainties of future surveys (like the LSST) with the same number and coordinates of SNe Ia as in the Union2.1 data set. Through this comparison, we showed that at low multipoles the constraining of a dipolar anisotropy can be improved up to about an order of magnitude for future surveys depending on the systematics.

In this work we considered only the lowest possible moment of anisotropy. But of course as the SNe Ia data sets increase in size and accuracy of observations, possible anisotropies in higher multipole moments i.e. at smaller angular scales could also be resolved in principle.

\section*{Acknowledgements}

HG would like to acknowledge  Seyed Mohammad Sadegh Movahed and Hajar Vakili to whom this project is indebted. SB acknowledges the hospitality of the Abdus Salam International Centre for Theoretical Physics (ICTP) during the final stage of this work.
Furthermore, we would like to express our gratitude towards Amir Hajian for all his unassuming help and guidance. We also benefited from illuminating discussions with Reza Rahimitabar. And finally, we thank Reza Ansari for his insightful discussions in the data analysis part.

\end{document}